# Quantitative phase imaging verification in large field-of-view lensless holographic microscopy via two-photon 3D printing

Emilia Wdowiak*, Mikołaj Rogalski, Piotr Arcab, Piotr Zdańkowski, Michał Józwik, and Maciej Trusiak**

*Warsaw University of Technology, Institute of Micromechanics and Photonics, 8 Sw. A. Boboli St., 02-525 Warsaw, Poland*

*emilia.wdowiak.dokt@pw.edu.pl*

**maciej.trusiak@pw.edu.pl*

**Abstract:** Large field-of-view (FOV) microscopic imaging (FOV > 100 mm$^2$) with high lateral resolution (1-2 μm for increased space-bandwidth product) plays a pivotal role in biomedicine and biophotonics, especially within the label-free regime, e.g., for whole slide tissue quantitative analysis and live cell culture imaging. In this context, lensless digital holographic microscopy (LDHM) holds substantial promise. However, one intriguing challenge has been the fidelity of computational quantitative phase imaging (QPI) with LDHM in large FOV setups. While photonic phantoms, 3D printed by two-photon polymerization (TPP), have facilitated calibration and verification in small FOV lens-based QPI systems, an equivalent evaluation for lensless techniques remains elusive, compounded by issues such as twin-image and beam distortions, particularly towards the detector's edges. To tackle this problem, we propose an application of TPP over large area to examine phase consistency in LDHM. In our research, we crafted widefield calibration phase test targets, fabricated them with galvo and piezo scanning, and scrutinized them under single-shot twin-image corrupted conditions and multi-frame iterative twin-image minimization scenarios. By displacing the structures toward the edges of the sensing area, we verified LDHM phase imaging errors across the entire field-of-view, showing less than 12% of phase value difference between investigated areas. Interestingly, our research revealed that the TPP technique, following LDHM and Linnik interferometry cross-verification, requires specific novel design considerations for successful large-area precise photonic manufacturing. Our work thus unveils important avenues toward the quantitative benchmarking of large FOV lensless phase imaging, advancing our mechanistic understanding of LDHM techniques and contributing to their further development and optimization of both phase imaging and fabrication.

## 1. Introduction

The capacity of optical microscopy to image transparent specimens has historically been a significant challenge, propelling the development and application of quantitative phase imaging (QPI) techniques.[1] Among various faces of QPI, Digital Holographic Microscopy (DHM) has gained considerable and well-deserved attention as an effective approach for quantitative measurements of phase distributions.[2] The forte of DHM lies in its ability to holographically process complex optical field information, encompassing both amplitude and phase modulations induced by specimen.[3] Notably, seminal in-line Gabor holography[4] realized digitally[5] forms the foundation of the lensless DHM (LDHM) framework,[6–8] appreciated for its hardware simplicity, cost-effectiveness, high-throughput and unrivaled large field-of-view (FOV easily surpassing 100 mm$^2$) with high space-bandwidth product. In the context of LDHM, holograms record the intensity of the optical field, which comprise two coherent terms, i.e., the object wave (+1 term) and the conjugate wave (-1 term), and the incoherent autocorrelation intensity term (0-th order). The twin-image effect is a well-known and studied major concern in LDHM (and generally on-axis DHM configurations) – a challenge initially tackled by Leith and Upatnieks[9] using off-axis architecture systems. Methods for lensless twin-image effect minimization include experimental approaches, e.g., phase-shifting[10,11] and off-axis configurations,[12,13] and numerical techniques, mainly iterative,[14] e.g., multi-wavelength,[15,16] multi-height,[17,18] regularization,[19,20] and deep learning.[21] Due to the high visual quality (large signal-to-noise ratio - SNR) of wide FOV imaging, LDHM ignited various applications,[22,23] especially in biomedicine for in-vitro live cell examination[24–28] and point-of-care scenarios.[29]

Although significant efforts have been implemented to improve the quality of LDHM imaging, a gap persists in the quantitative phase verification over large FOV. This work, therefore, proposes a new methodology to fulfill this need, implementing a new calibration phase test target design, precise manufacturing by specialized 3D printing via two-photon polymerization (TPP), and LDHM measurements to underscore the importance of twin-image removal and test its accuracy across entire FOV. Microscale phase structures, fabricated via TPP (Nanoscribe GmbH system), are already in place for lens-based high-magnification QPI systems,[30–32] showing great promise towards quantitative benchmarking of QPI technology in its various experimental and numerical implementations.[33,34] Recent works on this emerging topic consider on-the-fly monitoring of the structures manufacturing process.[35,36] To date, large FOV calibration targets have not been considered, which motivated this work to evaluate the LDHM-based high-throughput in-line QPI branch.

Taking advantage of newly designed and manufactured large-area phase targets, we study the sensitivity of LDHM for phase changes and attest to the precision of phase imaging across the entire sensing area, especially towards the edges. Our analysis also tackles an interesting issue, revealed after LDHM verification, that the 3D printing method itself necessitates a new design and manufacturing framework for effective large-area specialized fabrication. It is essential to mention that we also test the deployment of piezoelectric stages and galvanometric scanning and assess their role in facilitating the manufacturing process and generating related phase errors. The quantitative verification method proposed in this contribution offers exciting application prospects for precise diagnostic tools based on biomedical LDHM imaging, e.g., for early disease detection and drug discovery[24–29] through reliable large FOV phase monitoring of cells and tissue slices.

The manuscript is composed of four technical sections. In Section 2, we describe the experimental LDHM setup and numerical reconstruction algorithms used in our study. Section 3 is devoted to the presentation of the phase test design and manufacturing details. Section 4 analyses the LDHM phase imaging capabilities employing novel large-area phase targets, discusses the proposed methodology and obtained results. Section 5 concludes the manuscript.

## 2. Single-shot and multi-frame LDHM systems and algorithms used in the study

Figure 1 depicts the LDHM system used along the course of this study, composed of 405 nm laser diode (CNI Lasers, MDL-III-405-20 with FWHM = 23 pm) coupled with a single-mode fiber (Thorlabs, P1-460B-FC-1) and monochromatic camera (ALVIUM 1800 U-2050m) with Sony IMX183 CMOS sensor of 20.2MP (5496 (H) × 3672 (V)) with 2.4 x 2.4 $\mu m^2$ pixel and 116 $mm^2$ sensing area. A specimen is placed in between the source and camera planes. Semi-transparent sample placed on a microscope slide is illuminated with a coherent spherical wave coming from a quasi-point source of light (fiber's core). An in-line Gabor hologram (intensity image) is recorded in the detector plane formed as an interference of the wave scattered from the sample and an unscattered reference wave coming from the light source. To meet convenient holographic diffraction conditions, the sample should not be dense (cover too much of a radiation cone) to avoid excessive blocking of the reference beam[6] and excessive scattering. To retrieve both phase and amplitude information of the object located at a distance from the detector marked in Fig. 1 with $z$, a numerical backpropagation has to be performed. The detector-specimen to detector-source distances ratio determines the sample optical magnification in the measurement system. LDHM setup implemented in this study measures 300 mm in total (detector-source distance), while $z$ distance varies in the range of 1.8 – 4.5 mm, depending on the reconstruction method. Since the full setup length is significantly larger than $z$, magnification is considered constant, close to 1. Therefore, FOV in the setup is maximized and determined by the detector's sensing area. Numerical aperture (NA), dependent geometrically on sample-sensor distance and sensor size, varies in the range of NA 0.93 – 0.70 vertically and NA 0.96 – 0.83 horizontally, respectively, for 1.8 - 4.5 mm distances, nevertheless, being limited by detector's pixels size.

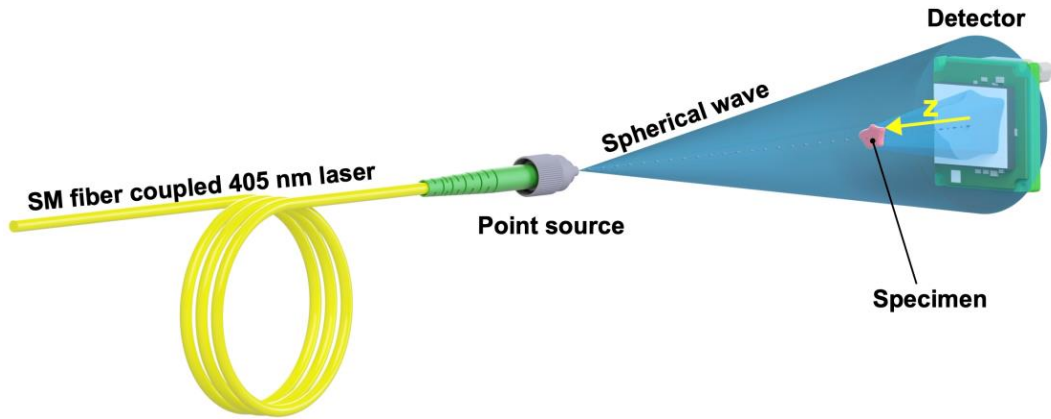

Fig. 1. The scheme of the LDHM system used with $z$ backpropagation distance indicated.

A straightforward numerical reconstruction method - angular spectrum (AS) backpropagation[37] - can be employed to retrieve the 2D complex optical field of the sample from a single hologram recorded on the chip. Since an in-line hologram is captured with only the intensity part of the complex optical field in the sensor plane, the twin-image artifacts disturb the reconstruction of the object plane. The twin-image effect manifests in the form of focused (object term) and defocused (conjugate term) information overlap, altering not only the reconstructed object's shape but also phase and amplitude quantitative accuracy. These errors impair post-process analysis of phase and amplitude maps, especially for complicated biological specimens (leading to, e.g., erroneous cell count). Across the variety of hardware and software solutions[10–19] implemented to minimize the twin-image effects, in this study, we investigate single-shot AS method, that is easy to perform in any setup, and robust iterative multi-height Gerchberg-Saxton (GS)[14] twin-image-correction approach employing data multiplexing.[17,18] The multi-height GS method requires capturing multiple hologram frames with different $z$ distances (in our case, achieved by moving the camera along the optical axis). Then, the iterative reconstruction algorithm is performed by propagating each hologram (starting from the closest to the object) to the subsequent hologram's plane, where the intensity constraints are imposed (obtained amplitudes are replaced with a square root of collected subsequent holograms, phases remain unchanged). Single iteration finishes after the propagation of the last (the furthest from the sample) hologram to the first hologram plane, and the algorithm is usually performed for 15-50 iterations (depending on the sample's characteristics and required reconstruction precision), after which the phases in the hologram planes are retrieved. In our system, we additionally employed complex field filtering constraints,[16] which enabled faster algorithm convergence and reduced the number of iterations to 5. We collected six holograms in 1.8 – 4.5 mm $z$ range, separated by approx. 0.5 mm. In the case of single-shot AS reconstruction, the minimal distance was kept between the camera and the specimen (around 1.8 mm, determined by the sample and camera dimensions).

It is important to emphasize that, for the primary method of numerical propagation between the camera and sample plane, we have employed the AS method, a well-established technique widely used for optical field propagation under the paraxial approximation, which applies to our specific case. To successfully reconstruct in-focus information using the AS method, it is essential to determine the distance between the sample and the camera accurately. For this purpose, we have utilized the DarkFocus[38] metric, a representative of autofocusing algorithms, which are of paramount significance in numerical focusing and play a crucial role in the LDHM. However, it is to be noted that, in situations involving larger magnifications and spherical wavefronts, appropriate propagation routines should be implemented. In a multi-plane scenario, it is important to determine the magnification of each hologram and correct the sample lateral shift due to the slight variation in magnification. It is performed via manual detection of characteristic features and affine transform-based hologram registration.[17]

## 3. Test target design and manufacturing

Two-photon polymerization can precisely fabricate elements exceeding Abbe's diffraction limit.[39] This is achievable due to the non-linear two-photon absorption phenomenon in the photoinitiators in a fabrication resin. To induce the absorption, fast-repetition of ultrashort (femtosecond) pulsed laser radiation is needed with the assistance of high NA optics to focus the beam and solidify an elementary volume of the material – a voxel.[39,40] To achieve high precision fabrication - dimensions below 200 x 200 nm$^2$ in the lateral plane and around 500 nm in the axial direction, usually, 63x or 100x magnification oil (or resin) immersion microscope objectives with NA 1.3-1.4 are used. In such conditions, the printing field is confined to approximately 200 µm diameter in galvanometric printing configuration and limited by the layer-by-layer movement implementation. For this reason, with the recent development of the TPP technique, multiple positioning approaches have been proposed to extend the accessible printing area.[41–43]

The most common method for additive fabrication with TPP is realized with the piezoelectric stage displacing the printing substrate, similar to macro-scale 3D printing.[39] However, such stages have a small travel range, usually around 300 x 300 x 300 µm$^3$. Fabrication of larger volumes with the use of the piezoelectric stages is associated with a significant increase in the manufacturing time (due to the 1000 µm/s speed limit in TPP Nanoscribe systems) along with the need for area stitching between subsequent printed fields. Such stitching can be realized with a lower precision mechanical stage. Unfortunately, it results in the creation of gaps, overlapping of the fields, and irregular shrinkage, decreasing the method's accuracy.[44] Alternatively, galvanometric scanning was introduced to control the beam's deflection while the substrate remains stationary.[45] This approach offers around two orders of magnitude faster fabrication[46] while theoretically keeping similarly high precision. Both approaches are available in the commercial TPP setup - Photonic Professional GT2 from Nanoscribe GmbH used in this study. Additionally, some approaches incorporate diffractive optical elements, micro-mirrors, or micro-lenses arrays to produce multiple printing beams (multiple foci) and parallel fabrication processes.[43,47] In this case, the fabrication is set to speed up significantly, constituting an efficient large-area solution for custom-assembled printing setups.

LDHM testing can significantly benefit from large-area two-photon printing. In principle, the test structure should meet (or even exceed) LDHM's resolution standards, at the same time preserving significant dimensions in the lateral plane to fully appreciate the method's space-bandwidth product. Taking advantage of both positioning approaches available in Photonic Professional GT2, a preliminary idea for the phase test target consisted of a matrix of small blocks divided into nine separated printing fields. The idle movement was performed with a high precision piezoelectric stage in the range of 300 x 300 µm$^2$, while fabrication movement was done within each of nine positions with fast galvanometric mirrors deflecting the beam in its 100 µm radius.[46,48] Therefore, given precise printing area reached up to 0.5 x 0.5 mm$^2$. Nevertheless, later analysis of geometrical and phase properties of the fabricated structure revealed, on the one hand, shape deviations of the blocks located towards the galvanometric deflection range end, and on the other hand, completely distorted phase distribution along the entire matrix (compared to designed constant value). Since the shape errors can be fixed by reducing the established galvo printing range, phase variation control is more challenging and might result from miscellaneous TPP aspects discussed below.

Phase change $\Delta\varphi$ introduced by 3D structure can be described with the following equation (1):

$$\Delta\varphi = k \cdot OPD = \frac{2\pi}{\lambda} \cdot h\Delta RI, \tag{1}$$

where $k$ – wave number is expressed in terms of wavelength $\lambda$ and $OPD$ – optical path difference, which is defined as a product of $h$ – geometrical axial thickness of a sample and $\Delta RI$ – refractive index difference between the object and surrounding medium. Particularly $h$ and $\Delta RI$ are structure-oriented variables possible to be controlled via two-photon fabrication method. The $\Delta RI$ depends on the chosen fabrication resin, while $h$ is limited by the TPP axial resolution. What is more, $h$ is prone to be influenced by the accuracy and repeatability of the substrate-resin interface localization method to be applied prior to the two-photon writing process, as well as any possible tilt deviation between the substrate and coordinate system of either piezo stages or galvo mirrors. Furthermore, there is a significant variation in substrates' thickness (from ±10 μm up to ±60 μm)[49,50], even for the ones that are specialized for two-photon fabrication. This aspect is especially noticeable when printing thin (< 3 μm) structures along larger distances (>100 μm), which we are innovatively interested in throughout this work.

To control the phase change within the structure's large volume, research and experimental tests were conducted, and a significant technological gap was found, which we proposed to fill via pre-printing a base layer unifying the axial coordinate system for the structure, Fig. 1. To achieve the highest test performance in LDHM conditions, the base layer must behave as a homogeneous background for the elements carrying phase information. Therefore, only a single printing field (no stitching required) performed with galvanometric scanning is taken into further consideration. To maximize the available galvo printing area, from this point in the study, only 25x, NA 0.8 immersion microscope objective is used with IP-S polymer resin (Nanoscribe GmbH). Using this hardware configuration, the theoretical resolution goes down to 600 nm laterally (which is still four times smaller than the resolution of our LDHM system) and 3300 nm axially, giving access to the 400 μm diameter galvo printing area.[51]

The final structure design is presented in Figs. 2(a) and 2(b) and verified in terms of its geometry in the Figs. 2(c)-2(e). Figure 2(c) consists of a scanning electron microscopy image of the test covered with a thin conductive gold layer (enhancing imaging conditions), validating the outer shape and layer-by-layer fabrication approach. Bright-field microscopy images included in the Figs. 2(d)-2(e) present the test's internal structure and its absorptive characteristics. The structure consists of three main components (marked in Fig. 2(a)): (1) – 6 μm thick base layer (orange in Fig. 2), (2) – phase resolution points (with height h of the same RI as base layer) carrying the phase information, and (3) – 7 μm thick solid immersion layer (blue in Fig. 2). Established thicknesses of (1)- and (3)-marked regions ensure mechanical resilience of the test. As we observed, with significantly decreased laser dose towards the edges of the 400 μm galvanometric printing area, the structure's total dimension is limited to 300 μm diameter. The base layer is chamfered to minimize phase discontinuity and diffraction artifacts (twin-image) appearance in later investigation after numerical holographic reconstruction. Four phase resolution points (2 - Fig. 2(a)) of 20 μm diameter are located in the central area of the base, also to minimize the twin-image effect both from the edges of the base and the presence of the other points. Lastly, the lower refractive index (RI) layer is printed around the resolution points to decrease the $\Delta RI$ value and more accurately control the object-background phase delay.

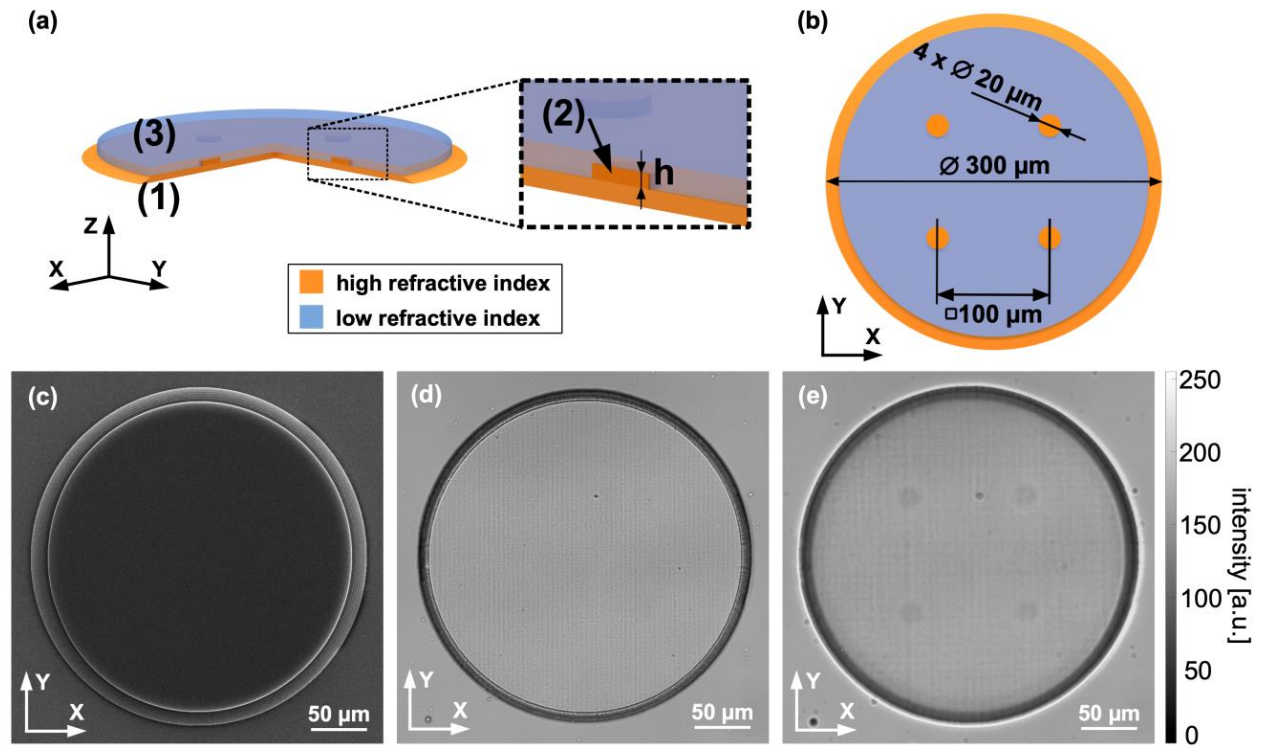

Fig. 2. Schematic representation of the phase test structure design – (a) isometric view with indicated regions: (1) – base layer, (2) – phase resolution points (h – axial thickness of a point), (3) – solid immersion; (b) top view with elements' dimensions. Verification of structure's geometry with (c) scanning electron microscopy, (d) bright-field microscopy – structure imaged in focus of upper surface, and (e) bright-field microscopy – structure imaged out of focus (visible phase resolution points).

Two-photon polymerization allows modulation of the RI characteristics of fabricated structure within the range offered by the used resin. The RI of a printed element relates to the monomer cross-linking in the material,[52] which can be altered by adjusting the laser dose parameters, such as writing laser power, printing density, scan speed, and aforementioned, writing (operational) movement approach – in our case galvanometric and piezoelectric. This way, the test target can vary in its optical phase delay characteristics, distinguishing the regions of higher ((1) and (2)) and lower (3) RI values (Figs. 2(a) and 2(b)). Since the TPP setup allows control over most of the factors modulating the structure's RI, changing just the writing movement approach alters the other parameters, such as scan speed limitations. Therefore, in this study, we rely on the general distinguishment between high and low RI characteristics in the structure and comparative study of LDHM large FOV abilities rather than attempting to estimate the theoretical phase delay value. According to product specifications,[53] the resin used in the fabrication process can vary in RI value in the range of 1.486-1.515 ($\Delta$RI = 0.029, from liquid to fully polymerized form of resin). Therefore, the difference in RI values between fully and partially polymerized regions (higher and lower RI) within the structure can achieve about 0.02.

To fabricate a uniform base layer in a reasonable time, a fully galvanometric scanning-positioning approach was adopted. Nevertheless, fluctuations of laser energy dose have been observed within the aperture of the printing area. Figure 3(a) presents the reference measurement of the test structure printed with the galvanometric scanner. The optical evaluation of fabricated structures was carried out with a Linnik interferometer[54] employing a 5-frame temporal phase shifting phase reconstruction algorithm. To achieve high qualiy results, the interferometer consists of an LED light source of 635 nm wavelength. For comparison purposes, a similar structure was fabricated with exclusive use of the piezoelectric fabrication movement, see Fig. 3(b). The piezoelectric-printed structure reveals diagonal tilt, still being much more uniform than the galvo-printed one containing vertical errors, marked in Fig. 3(a) with a red arrow. However, fabrication time starts to play a noticeable role – printing this particular structure with galvo took around 30 minutes, while piezo accomplished the same body in around 9 hours of a constant print. To broaden further LDHM analysis, piezo and galvo-positioned structures are investigated in this study. The errors present in the galvo printing aperture are, certainly, an undesirable effect but can bring a new perspective to LDHM imaging in the presence of even more challenging test structures in comparison with the piezo printing efficiency.

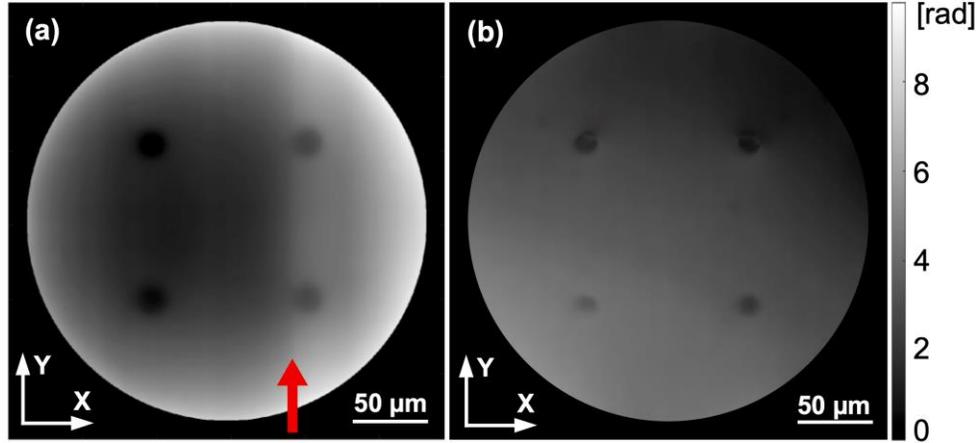

Fig. 3. Phase maps of the test structures fabricated with: (a) galvanometric scanning and (b) piezoelectric stages.

All structures used in the following parts of this study are fabricated with IP-S polymer resin. Tests are printed on 25 x 25 x 0.7 mm$^3$ ITO-coated soda lime glass substrate. Before the printing, the glass substrate is prepared in a silanization process to enhance the structure's adhesion to the substrate.[55] After the TPP process, the structure is cleaned out of unpolymerized liquid resin in 12 minutes of isopropyl alcohol bath and then dried, completing the fabrication process.

## 4. Experimental evaluation of the LDHM phase imaging capabilities

Considering fabrication and design constraints, as well as accommodating their use to the large FOV LDHM testing application, the following novel methodology is proposed. Figure 4(a) shows, to scale, a schematic representation of the entire FOV available in the used LDHM setup (Fig. 1). Four regions are marked with different colors: (1) upper side, (2) corner position, (3) central position, and (4) right side. Each one consists of three custom-designed phase test targets, enlarged in a box in Fig. 4(b). Within these three tests, the phase change is modulated by varying geometrical thickness of the resolution points (visible in A-A, B-B, and C-C cross-sections in Fig. 4(c)). A-A resolution points reach 2 μm thickness, which increases with a 0.5 μm relative difference between A-A and B-B, and B-B and C-C, while all the other fabrication parameters remain constant. Thus, structure A-A is associated with low, B-B with medium, and C-C with high phase change within the body. Including the base layer in the design allows to fabricate axial dimensions that are smaller than theoretical 3.3 μm resolution in Z-axis. In such case, resolution points' thickness is established in relation to the base layer level. Theoretically, C-C points representing high phase change should not exceed 2π radians in their phase delay value, due to the presence of lower RI surrounding layer (small ΔRI leading to Δ$\varphi$ < 2π according to equation (1)). Nevertheless, resolution points are still prone to be influenced by the operational movement approach and other factors determining RI in TPP fabrication, described in Section 3. Four positions (Fig. 4(a)), in which test triplets can be found, indicate borderline displacements, where the triplet has been moved to study LDHM performance in potentially most challenging and neuralgic positions within the camera sensor area.[5–8,27]

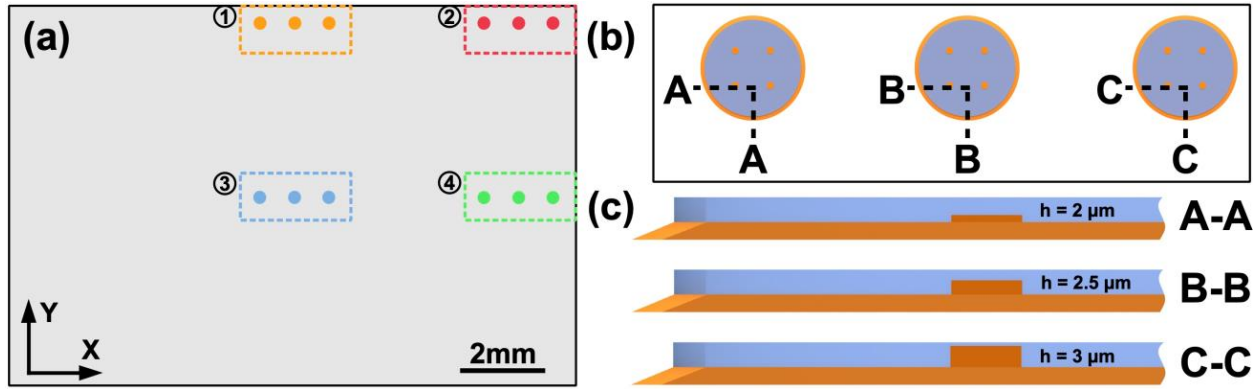

Fig. 4. Indication of four testing spatial locations and three types of phase-sensitive elements: (a) full accessible FOV (areas (1) – (4) indicating measured borderline positions), (b) test triplet closeup, (c) tests' cross-sections indicating varying thickness of the resolution points (not to scale).

Both piezoelectric and galvanometric control were used in the TPP process. In this way, two sets of structures (each containing three elements) were fabricated – piezo and galvo triplets. Each set was moved within the camera FOV to be measured in four distinctive representative positions, described in the previous paragraph. First, we present the phase imaging results for a single-shot AS and multi-shot GS reconstructions in Fig. 5. Piezo- and galvo-printed triplets' phase maps are indicated with colors and numbers (1) – (4), according to the camera sensor measuring positions, introduced in Fig. 4. We can observe characteristic twin-image errors in all single-shot reconstructions. These errors are amplified by the presence of the base and solid immersion layers and sharp $\Delta RI$ jumps on the borders of these two, as well as the $\Delta RI$ value in reference to the printing substrate. On top of that, in galvanometric print, one can notice additional vertical errors (marked in in Fig. 3 and enlarged in Fig. 5) resulting from the galvo-scanning-related instability of the induced polymerization process. Figure 5 presents the piezo process with significantly limited printing errors via the manufacturing process only. In multi-frame GS iterative reconstructions, we can observe characteristic circular twin-image error suppression in both piezo and galvo reconstructions. Fabrication process errors can be observed similarly to the single-shot reconstructions, as they are not numerically induced. Some phase maps presented in Fig. 5 (e.g., piezoelectric print AS (4)) display low-contrast fringe pattern artifacts in the background of the structures. This results from a random interference caused by slight non-parallelism of planes of the substrate and the detector in considered measurements.

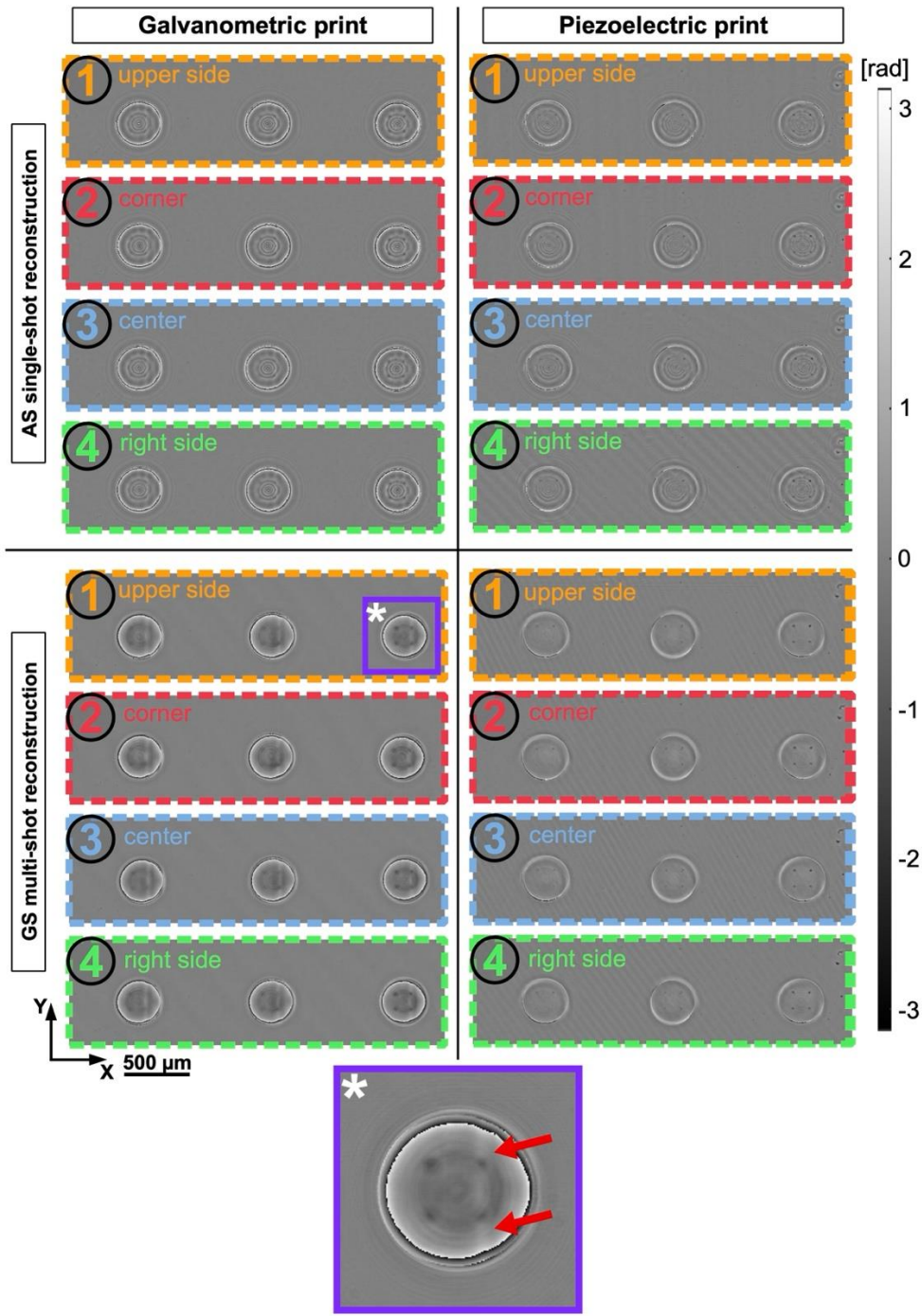

Fig. 5. Phase maps reconstructed with single-shot AS method and multi-height GS approach: piezo-printed and galvo-printed tests. Regions indicated with numbers (1)-(4) are associated with different FOV positions: (1) – upper side, (2) – corner, (3) – central, (4) – right side. Marked colors and numbers agree with the methodology indicated in Fig. 4. Region marked with * enlarged to expose fabrication errors.

For a more detailed quantitative evaluation of single- and multi-shot phase imaging sensitivity and reliability, we present, in Fig. 6, cross-sections through low, medium, and high phase change elements in the central position of the detector's sensing area. Additionally, all graphs in Fig. 6 include references to the Linnik interferometry measurement of the structures. Reference cross-sections have reduced low-frequency trends (resulting from aberrations and tilt, visible in Fig. 3) for better LDHM comparison. Interestingly, LDHM inspection seems to lose that information, revealing only high-frequency resolution points and noise. This is related to the issues with retrieving low frequencies in the in-line and non-interferometric QPI setups, acknowledged in the literature.[20]

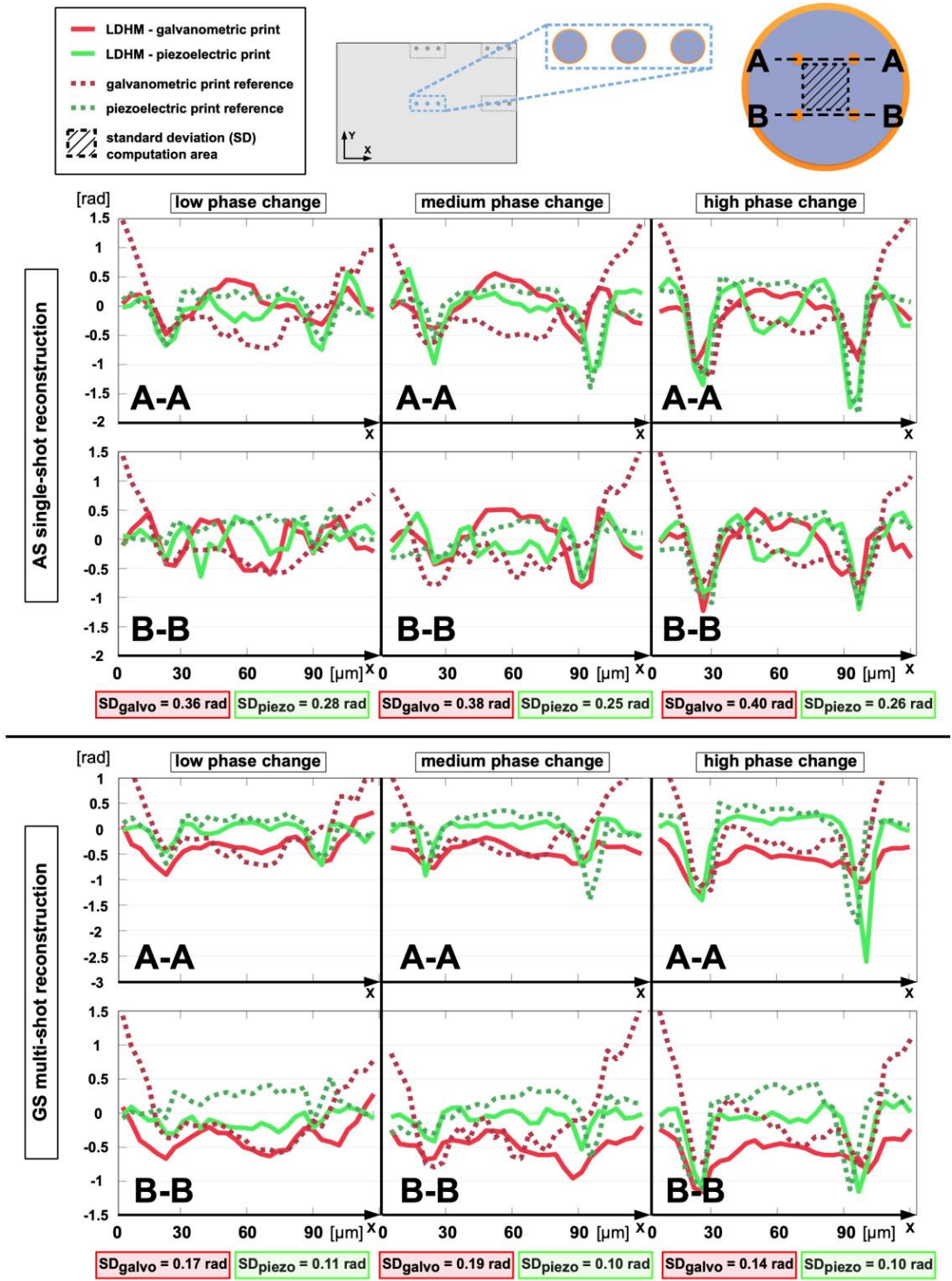

Fig. 6. Cross-sections of LDHM-measured low, medium, and high phase change resolution points in the central area of the FOV, reconstructed with single-shot AS method and multi-height GS approach. Cross-sections are analyzed for galvanometric and piezoelectric print along A-A and B-B lines, marked on the schematic test target. A Linnik interferometer-measured reference is included for each case and is indicated with dotted lines. Standard deviation was calculated for LDHM-measured structures for the area marked in the schematic structure.

Single-shot approach cross-sections show a significant noise level, confirmed with standard deviation calculated for the area between resolution points, marked in Fig. 6 schematic representation of a structure. Especially low phase changes, printed with both piezo and galvo approaches, exhibit very poor phase delay signal-to-noise ratio, making it almost impossible to distinguish the phase resolution points. Medium and high phase resolution points are noticeably better reconstructed; however, they still indicate prominent twin-image disturbances. Multi-height results demonstrate substantially lower noise levels, making it possible to distinguish even the smallest phase changes in piezoelectric print. Standard deviation, compared to the results from single-shot reconstruction, is reduced by an average of 60% in piezo print and 57% in galvo print. What is more, values in piezo-controlled base layers oscillate around 0 radians, validating the correct functionality of the specialized base for precise large-area TPP. At the same time, galvo bases introduce additionally around -0.5 rad phase delay to each resolution point, distributed non-uniformly through the layer due to printing instability-related effects.

For more quantitative insight into the large FOV investigation, Fig. 7 includes cross-sections through the structures representing the most distinct phase delay (because of the highest SNR) - third structures from the triplets. The graphs gather cross-sections from all four areas in the FOV for piezoelectric and galvanometric printing reconstructed with single-shot and multi-frame methods, with Linnik interferometry (Fig. 3) measurements for reference.

Plots included in Fig. 7 show high compliance between all cross-sections in analyzed regions. Variation between the lines (positions in the FOV) in single-shot reconstructed graphs results from overall higher noise level (and twin-image artifacts) in the AS method reconstructed phase maps. There is no apparent pattern indicating the dependence of the phase value and geometrical shape of a structure on the position in the sensor area. Those minor variations presumably come from measurement noise. Even the most challenging conditions, such as galvo-printed structures reconstructed with the single-shot method, demonstrate relatively high phase profile agreement.

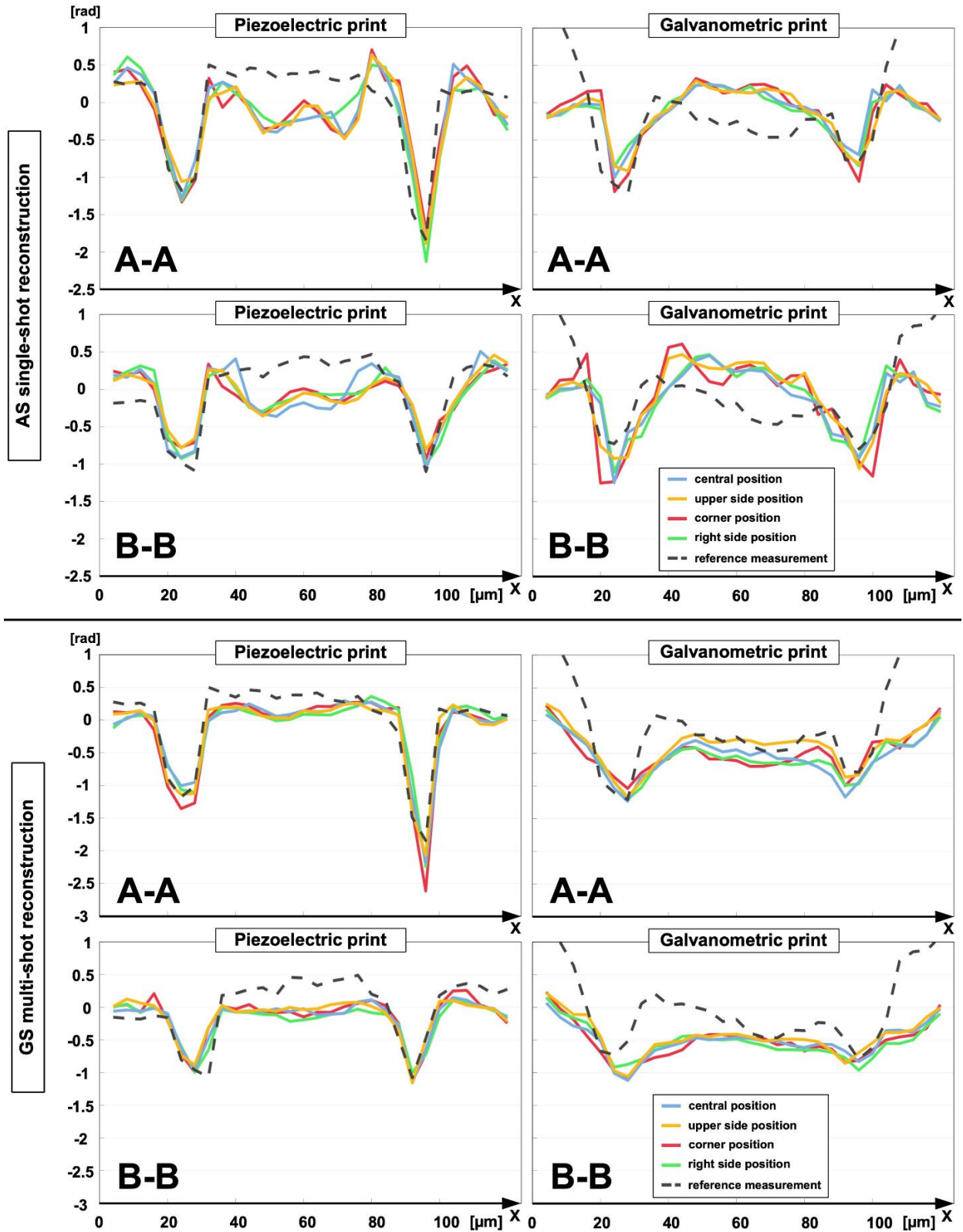

Fig. 7. Cross-sections along the lines A-A and B-B indicated in the scheme from Fig. 6 of the test targets introducing the highest phase delay, located in central, upper side, corner, and right side positions in the FOV of the detector. The data was reconstructed with single-shot AS and multi-shot GS methods, for piezoelectric and galvanometric prints.

Table 1 introduces a quantitative analysis of the results. Averaged phase values are calculated (in radians) for only one (lower left) resolution point – the same element is chosen for all analyzed structures (third structures in a triplet row, carrying the highest phase change). The results in Table 1 are expressed as a phase difference between the value averaged from 5 pixels containing maximum values from the resolution points areas and the median value of the area surrounding points within the test (Fig. 8). The analyzed 5-px region is smaller than the theoretical size of the resolution point (marked on Fig. 8), to deliver the most realistic phase value, not influenced by the twin-image and points' shape deviations. In terms of detector pixels, the single object size should involve a circular area of around 8 px diameter (Fig. 8). As we perform LDHM imaging in significantly large FOV, each resolution point is sampled with a very small number of pixels. Moreover, due to manufacturing and reconstruction imperfections, resolution points' shapes are deviated. Therefore, analyzing mean value across a fixed circular area would contribute to mixing the point's phase information with the surrounding area values, providing a delusive outcome (Fig. 8). Our 5-px average methodology emphasizes that the primary aim of this work is the comparative quantitative study of large FOV LDHM phase imaging, within on-axis conditions, rather than reconstruction or manufacturing analysis itself.

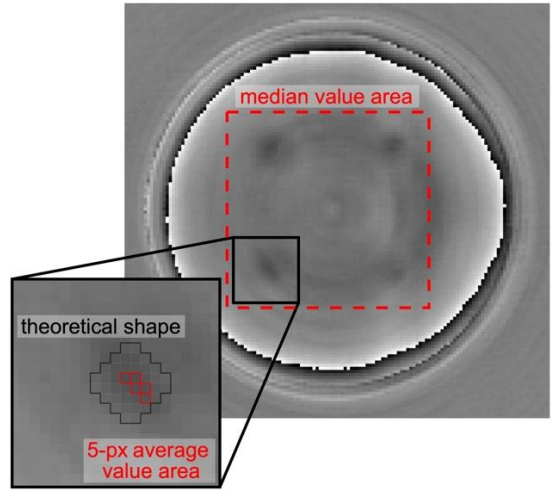

Fig. 8. Visualization of the regions analyzed quantitatively in the study: background area from which median value is calculated (marked with dotted line rectangle) and chosen 5 pixels containing maximum values within theoretical resolution point area (marked with red squares).

Moreover, Table 1 includes the phase change ratio comparing averaged values from non-central positions to the central one (nomenclature as in Fig. 4), according to equation (2).

$$Center\ value\ ratio = (1 - \frac{Average\ phase\ [rad]}{Average\ central\ phase\ [rad]}) * 100\% \tag{2}$$

The closer is *the center value ratio* to 0%, the more similar the measured edge value is to the center – the higher center-edge phase compliance. A positive *center value ratio* indicates that the phase measured in the center position achieves a higher value than the one measured on the edge. Table 1 also indicates reference measurement of analyzed tests (shown in Fig. 3), where the average phase value is computed similarly to the LDHM results. In this case, the analyzed region is sampled significantly better (due to the use of high-magnification imaging optics); therefore, the mean is calculated from 20 pixels of maximum values. What is more, comparing sampling conditions between the two methods, it is worth noticing that low sampling of LDHM exposes the technique to the influence of the high-frequency noise disturbances and the TPP fabrication inaccuracies, thus promoting heterogeneity of the entire manufacturing-to-imaging process.

Tab. 1. Quantitative evaluation of large FOV LDHM for single-shot AS and multi-height GS reconstruction methods of lower left resolution point (in the third test structure with highest designed phase delay).

| **Single-shot AS method reconstruction** | | | | | | | | |
|---|---|---|---|---|---|---|---|---|
| | **Reference method** | **Central position** | **Upper side position** | | **Right side position** | | **Corner position** | |
| | Average reference value [rad] | Average phase [rad] | Average phase [rad] | Center value ratio [%] | Average phase [rad] | Center value ratio [%] | Average phase [rad] | Center value ratio [%] |
| Piezoelectric print | **-0.8653** | **-0.9894** | **-0.8190** | 17.22 | **-0.9348** | 5.52 | **-0.8902** | 10.03 |
| Galvanometric print | **-1.4293** | **-1.2324** | **-1.0654** | 13.55 | **-1.1603** | 5.85 | **-1.5041** | -22.05 |
| **Multi-height iterative approach reconstruction** | | | | | | | | |
| | **Reference method** | **Central position** | **Upper side position** | | **Right side position** | | **Corner position** | |
| | Average reference value [rad] | Average phase [rad] | Average phase [rad] | Center value ratio [%] | Average phase [rad] | Center value ratio [%] | Average phase [rad] | Center value ratio [%] |
| Piezoelectric print | **-0.8653** | **-1.0392** | **-1.0682** | -2.79 | **-1.1237** | -8.13 | **-1.1573** | -11.36 |
| Galvanometric print | **-1.4293** | **-0.7491** | **-0.7434** | 0.76 | **-0.6204** | 17.18 | **-0.6768** | 9.65 |

The results included in the Tab. 1 confirm the trends noticed throughout the Fig. 7. What is more, *the center phase value ratio* does not exceed the absolute value of 22.05% change in the least favorable conditions (galvanometric print in corner position), mostly varying in the range of ~10%. Once again, there are no apparent trends related to the edge FOV position, even considering the rectangular shape of the detector matrix (3:2 proportion, inducing spatially non-uniform sampling of the holographic shadow coding the object information). Data covered in Tab. 1 quantitatively verify the successful twin-image reduction of the multi-height reconstruction over the single-shot method (smaller *center value ratio*). Notably, multi-height reconstruction applied to the piezo-printed structures greatly reduces the large FOV errors, making this case the most consistent result along all the analyzed conditions. Nevertheless, it is worth noticing that, unlike others, these results give negative *center value ratios* (edge average phase is larger than center value). Galvanometric print also benefits from the multi-shot approach, especially considering high variations of center value ratio across full FOV in the AS single-shot method (almost 35% of the absolute difference between the upper side and the corner). The absolute variation of the multi-height iterative method reaches just 16.42% for galvo-printed structures. Presented numbers also highlight that LDHM, due to large FOV and small magnifications (poor sampling of small features) combined with measurement noises, gives quantitative phase results within ± 25% of the expected value (around 1 rad), which is the first, to the best of our knowledge, assessment, and recommendation of its use in large FOV single-shot quantitative phase imaging. The selected multi-shot approach improves overall consistency of the quantitative character of LDHM phase imaging, at the expense of the limited temporal resolution (multi-shot hologram recording).

Further analysis, in Fig. 9, includes four bar charts of phase values (analyzed according to Fig. 8) evaluated across each resolution point (upper and lower, left and right) in the considered structure. Graphs also incorporate the variance values calculated for every reconstruction method in reference to the point's FOV position. Notably, Fig. 9 reveals significant inconsistency in phase value between all four resolution points within the structure. As we express phase value in the form of the phase difference between the point and its surrounding area, the heterogeneity should be more prominent mostly in the structures with galvo-printed base and solid immersion layers (varying RI character). However, different phase levels of four points manifest in both piezo and galvo structures, independently of the reconstruction method, implying TPP fabrication inaccuracy. Nevertheless, Fig. 9 validates conclusions drawn on the detailed evaluation performed in Tab. 1, showing no trends associating phase accuracy with large FOV sample localization. The variance oscillates mostly around the order of magnitude $10^{-2}$–$10^{-3}$ radians, suggesting low data fluctuations. Since piezo-printed structures achieve phase results of similar orders of magnitude for both reconstruction approaches, when it comes to the galvo-printed structures, the multi-height method seems to reduce twin-image effects at the expense of lowering the overall mean phase value in the entire FOV.

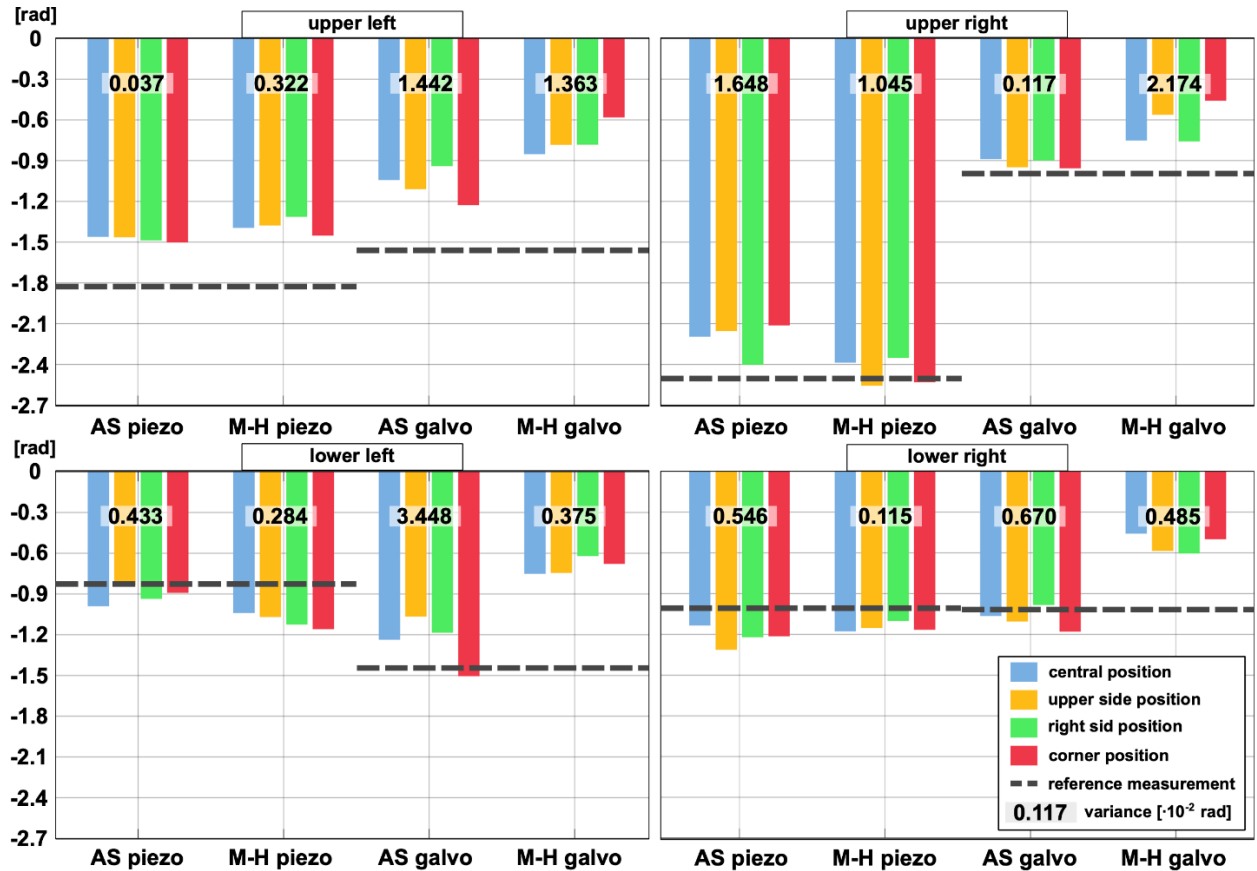

Fig. 9. Bar graph evaluation of quantitative study of large FOV LDHM for single-shot AS and multi-height methods of reconstruction of four resolution points within a single structure (far right structure in the triplet).

Large FOV LDHM imaging, analyzed in experimental evaluation, demonstrates solid agreement between the central detector position and three significantly distant positions. Deviations of obtained phase values are influenced mainly by the twin-image artifacts and random noise variables combined with the low sampling rate of small phase features in the low magnification scenario studied here. Nevertheless, the methodology by the of performed evaluation, although significantly improved proposed base layer deployment, is subject to errors of large area TPP fabrication technique (within a single test element containing four resolution points). While overall shape limitations (Section 3) are constraining the test's performance, also the spatio-temporal accuracy of the fabrication process starts to play an important role.

Considering the quality of fabricated structures, one can notice a common tilt of the set of three structures (subtle deviation in the Z-axis position), resulting in uneven phase distribution along four resolution points. Both galvo and piezo print fabrication of the base layer did not entirely reduce the influence of coordinate systems' inconsistency/significant area printing issues. Therefore, the large FOV evaluation has to be determined by an individual point compared in varying positions due to TPP manufacturing shortcomings. It does not limit the quantitativeness of the performed study but generally questions the large area TPP purpose. What is more, Figs. 6 and 7 demonstrate the repercussions of the design and fabrication divergence. The differences are verified with the reference method; however, galvo-printed tests are still affected by non-linear errors within large area aperture in a single calibration structure. For this reason, piezoelectric printing seems to be the significantly more metrologically-relevant approach for large-area TPP fabrication, even considering its very high manufacturing time.

We presented that by applying an iterative multi-height reconstruction approach to the considered data, the twin-image level is significantly reduced in all FOV position scenarios (Fig. 6 and 7). This reconstruction procedure does not affect the quantitativeness of the performed study compared to the single-frame AS method and the reference study. What is more, according to Tab. 1, multi-shot reconstruction improves phase homogeneity by comparing edge camera sensor measurements to the center position ones. This validates the particular multi-shot iterative reconstruction method in quantitative phase study in large FOV. An individual evaluation of each piezo and galvo set of the structures gives an opportunity not only to identify small phase values detection and LDHM reconstruction but also to benchmark further large FOV evaluation domain (one central and three edge positions) in terms of two considered reconstruction methods (single-shot and multi-frame approaches). In this manner, the proposed methodology establishes a relevant tool for the quantitative comparative study bridging large FOV on-axis and small FOV off-axis holographic phase imaging.

## 5. Conclusions

The rapidly advancing field of optical microscopy, particularly in the domain of lensless digital holographic microscopy, has presented both challenges and opportunities. This research addresses a key issue in LDHM - the need for more quantitative verification of large field-of-view imaging - and, in doing so, has made significant strides toward a better understanding of LDHM and its more precise and reliable applications. Our novel use of two-photon polymerization 3D printing (nonlinear photonic fabrication across a large area) has facilitated the first systematic study of phase reconstruction consistency in large FOV LDHM imaging. The design and manufacture of innovative large-area calibration phase test targets have allowed us to quantitatively and qualitatively examine the robustness of LDHM to twin-image effect under both single-shot and multi-frame scenarios. Our experimental work has highlighted the importance of twin-image removal in LDHM and showcased the need for enhanced sensitivity to small phase changes. Critically, we demonstrated the precision of LDHM phase imaging across the entire FOV, even towards the edges, a feature that can have important implications for the accuracy and utility of LDHM, especially in biomedical applications, e.g., high-throughput examination of full-colony of live cells. The phase error concerning the central FOV position imaging does not exceed 23% of the *central value ratio* across four studied resolution points, two fabrication methods (piezoelectric and galvanometric control), and two distinct reconstruction approaches. Moreover, when performing in preferable conditions (piezo printed structure, reconstructed with an iterative algorithm), center-edge phase error did not exceed 12%. Notably, we corroborated that the 3D printing via a two-photon polymerization approach necessitates special design considerations for successful large-area manufacturing, as evidenced by our recommendation of piezoelectric stages over traditional galvanometric scanning for better quality outcomes. Nevertheless, errors that appear in large areas of galvo and piezo printing should be considered in further studies. We showed that both two-photon printing and LDHM phase reconstruction are rather heterogeneous methods and need further improvements (preferably some correction mechanisms) in viability. Our findings point to exciting new pathways for quantitative benchmarking of large FOV phase imaging, possibly bridging a significant gap in the current understanding and application of LDHM. This work has broader impacts on the fields of high-throughput biomedical imaging, early disease detection, drug development, and cell physiology and pathology diagnostics, where large FOV LDHM is increasingly employed.

In conclusion, our research has not only addressed the pressing need for large-area quantitative verification of LDHM but also paved the way for further advancements in the field. We anticipate that the methodologies and insights gleaned from this work will catalyze further refinement in LDHM and contribute significantly to its future trajectory.

**Funding.** Funded by the European Union (ERC, NaNoLens, Project 101117392). Views and opinions expressed are however those of the author(s) only and do not necessarily reflect those of the European Union or European Research Council Executive Agency (ERCEA). Neither the European Union nor the granting authority can be held responsible for them.

This work has been funded by the YOUNG PW grant financed by Warsaw University of Technology through the Excellence Initiative: Research University program (Ministry of Education and Science, Poland). The research was conducted on devices cofounded by the Warsaw University of Technology within the Excellence Initiative: Research University (IDUB) program. M.R. is supported by the Foundation for Polish Science (FNP start program).

**Disclosures.** The authors declare no conflicts of interest.

**Data availability.** Data underlying the results presented in this paper are not publicly available but may be obtained from the authors upon reasonable request.